# PHYSICAL MODELING OF AIRFLOW-WALLS INTERACTIONS TO UNDERSTAND THE SLEEP APNEA SYNDROME

Y. Payan[1], X. Pelorson[2] and P. Perrier[2]

(1) TIMC/GMCAO, Institut Albert Bonniot, UMR CNRS 5525 & Université Joseph Fourier, Domaine de la Merci, 38706 La Tronche Cedex, France. Yohan.Payan@imag.fr

(2) Institut de la Communication Parlée, INPG-Univ. Stendhal, UMR CNRS Q5009, 46 Ave. Félix Viallet, F-38031 Grenoble Cedex, France.

## 1. ABSTRACT

Sleep Apnea Syndrome (SAS) is defined as a partial or total closure of the patient upper airways during sleep. The term "collapsus" (or collapse) is used to describe this closure. From a fluid mechanical point of view, this collapse can be understood as a spectacular example of fluid-walls interaction. Indeed, the upper airways are delimited in their largest part by soft tissues having different geometrical and mechanical properties: velum, tongue and pharyngeal walls. Airway closure during SAS comes from the interaction between these soft tissues and the inspiratory flow.

The aim of this work is to understand the physical phenomena at the origin of the collapsus and the metamorphosis in inspiratory flow pattern that has been reported during SAS. Indeed, a full comprehension of the physical conditions allowing this phenomenon is a prerequisite to be able to help in the planning of the surgical gesture that can be prescribed for the patients.

The work presented here focuses on a simple model of fluid-walls interactions. The equations governing the airflow inside a constriction are coupled with a Finite Element (FE) biomechanical model of the velum. The geometries of this model is extracted from a single midsagittal radiography of a patient.

The velar deformations induced by airflow interactions are computed, presented, discussed and compared to measurements collected onto an experimental setup.

## 2. INTRODUCTION

Sleep apnea is a disorder in which a person stops breathing during the night, usually for periods of 10 seconds or longer. In most cases the person is unaware of it. This disorder results from collapse and obstruction of the throat pharyngeal airway (figure 1, left). It is accepted that this occurs due to both a structurally small upper airway and a loss of muscle tone, as there is loss of the wakefulness stimulus to upper airway muscles with sleep onset. This results in airway collapse, increased resistance to airflow, decreased breathing, and increased breathing effort [1]. In most subjects, the narrowest airway cross-section area occurs behind the palate and uvula. This area is the most vulnerable to obstruction from loss of muscle tone during sleep. On physical exam, a long and wide velum, large tonsils, and redundancy of pharyngeal walls may be found. Lower throat findings may include a large tongue and lingual tonsils.

An estimated 5 in 100 people, typically overweight middle-aged men, suffer from sleep apnea. In addition to a strong daily fatigue, several chronic cardiovascular complications have been found to be related to sleep apnea syndrome (SAS), such as systemic arterial and pulmonary hypertension, heart failure or arrhythmias. The Apnea Index (AI) measures the number of apneas per hour. Hypopnea is defined as a decrease in airflow of 50% or more (without complete cessation) accompanied by a drop in oxygen saturation. The Hypopnea Index (HI) is the number of hypopneas per hour. The Apnea/Hypopnea Index (AHI) is the sum of AI and HI. The definition of sleep apnea in terms of the AHI is assumed to be the following: normal is AHI 0-5, mild 5-15, moderate 15-30, and severe 30+.

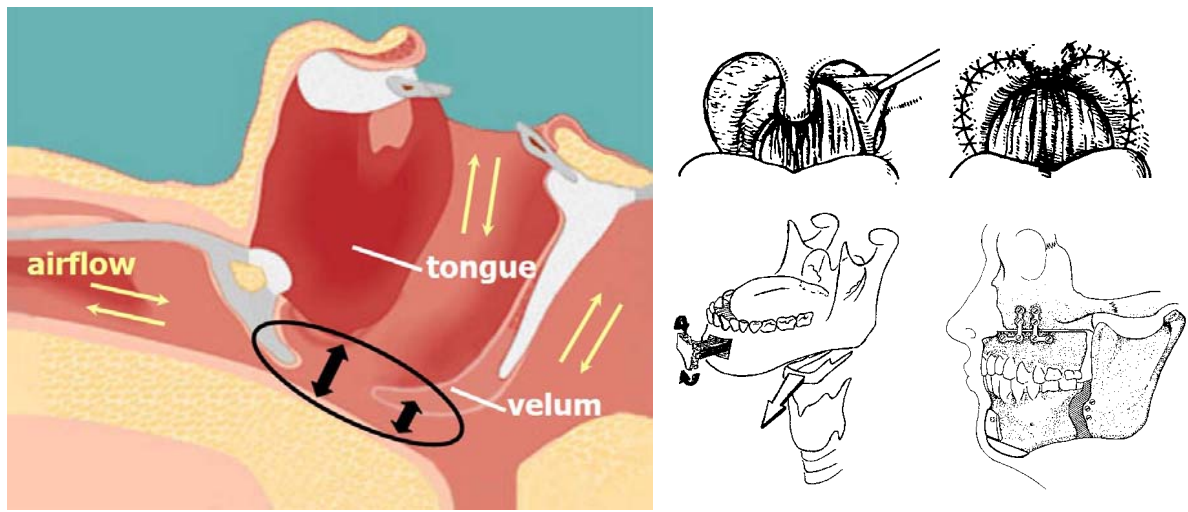

Fig. 1: Sleep Apnea Syndrome: fluid (airflaw) / wall interactions (left) and surgical treatment (right).

Surgical techniques used for the treatment of the SAS can either reduce the volume of the tongue, stiffen the velum, or try to have a more global and progressive action on the entire upper airways (figure 1, right). In order to fully understand the metamorphosis occurring in the inspiratory flow pattern that has been reported during SAS, some mechanical models of the upper airway have been developed, assuming that upper airways can be represented by a single compliant segment ([2], [3]), or by series of individual segments representing singularities [4]. In this aim, a complete biomechanical model of the upper airways appears thus to be interesting, to describe and explain — at the physical point of view — the upper airway obstruction.

The work presented here focuses on a simple model of fluid-walls interactions. A 2D Finite Element (FE) biomechanical model of the velum is introduced, and coupled with an analytical approximation of the equations that govern the airflow inside a constriction. The simulations of airflow / walls interactions are then compared with *in vitro* measurements collected onto an experimental setup.

3. FINITE ELEMENT MODEL OF THE VELUM

Lingual, velar and pharyngeal soft tissues are partly responsible for the SAS as their deformations can even lead to a total closure of the upper airways. In a first step, we only focus on the velo-pharyngeal region, at the intersection between the nasal and the oral cavity. In this perspective, a 2D sagittal continuous model of the velum was elaborated and discretized through the Finite Element Method. The codes developed for this model assume no displacement in the transverse direction (*the plane strain*

*hypothesis*) as well as a small deformation hypothesis. Velar tissues being assumed as quasi-incompressible (because mainly composed of water), a value close to 0.5 was chosen for the Poisson ratio. A 10 kPa value was taken for the Young modulus, which seems coherent with values reported for tongue [5] and vocal folds [6]. The geometry of the model was extracted from a single midsagittal radiography of a patient (figure 2).

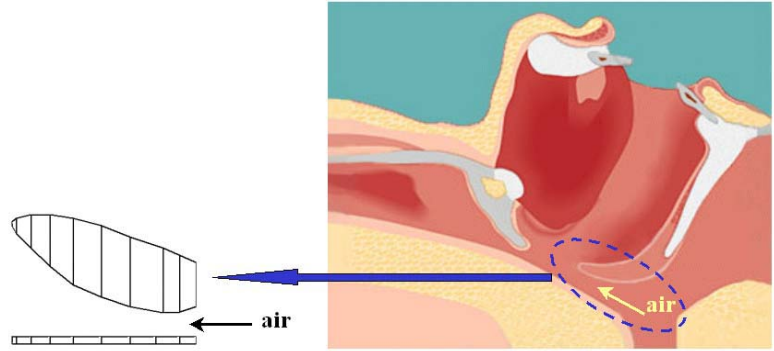

Fig. 2: Midsagittal view of the upper airways (right) and FE model of the velum (left).

The upper part of the model represents the velar tissues. Geometrical dimensions were close to values reported in the literature [7] [8]: the total length is of order of 30-mm while thickness varies from one extremity to the other (with a mean of 5-mm).
Only the velum deformations were taken into account for the current simulations. The pharyngeal walls (the lower part of the model in figure 2) were assumed to be rigid. The two points located onto the right part of the velum were also considered as fixed in order to model the velar attachment to the hard palate. Finally, simulations were limited to the 2D midsagittal plane, but for the computation of pressure forces (integrated along the 3D geometry), a 30-mm value was taken for the velar thickness in the frontal plane.

## 4. PHYSICAL MODELING OF THE AIRFLOW

From a fluid mechanical point of view, the partial or the total collapse of the upper airway, as observed during sleep hypopnea or apnea, can be understood as a spectacular example of fluid-walls interaction. While the most important parameters influencing this effect *in vivo* are well known, this phenomenon is still difficult to model and thus to predict. Figure 3 presents in a simple way a constriction inside the upper airways.

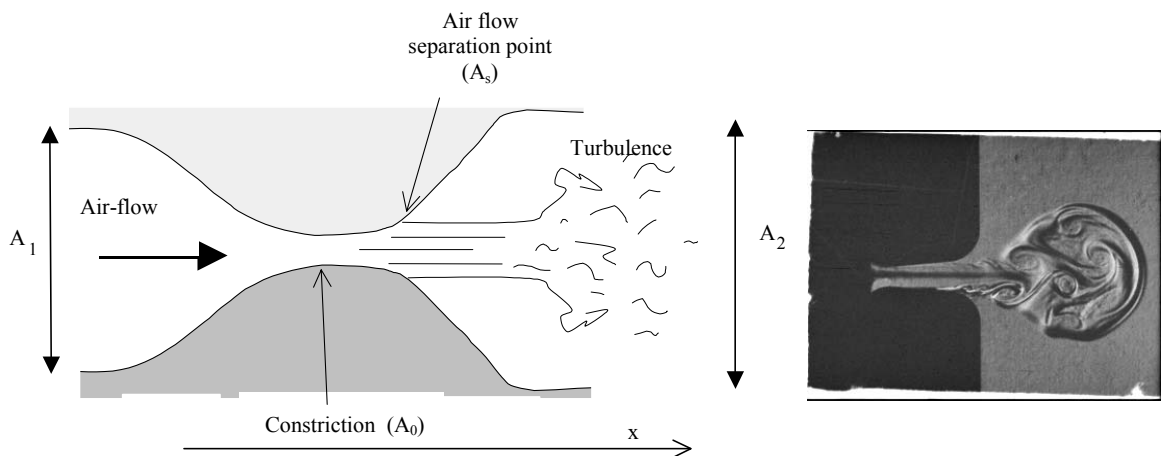

Fig. 3: Schematic Illustration for airflow inside the constriction.

An exact analytical solution for the flow through such a constriction is not available. Further, full numerical simulations of the unsteady three-dimensional flow through a deformable structure are still, at present time, very complex even using the recent numerical codes and using powerful computers [9]. For these reasons, and also because the aim of this paper is to provide a qualitative description of a sleep apnea, we use here a simplified flow theory based on the following assumptions:
- As the airflow velocity in the upper airways is, in general, much smaller than the speed of sound, it can be assumed that the flow is locally incompressible.
- It can be reasonably assumed that the time needed for the constriction to collapse (of order of a second) is large compared with typical flow convection times (the time needed for the flow to pass the constriction is of order of a few milliseconds). Therefore, it will be assumed that the flow is quasi-steady.

The principle of mass-conservation thus yields the following relationship:

$$\Phi = \text{constant} \qquad (1)$$

where $\Phi = v.A$ is the volume flow velocity, $v$ and $A$ are respectively the (local) flow velocity and upper-airways area. As a third assumption, it is considered that all viscous effects can be neglected. This assumption can be rationalized partially by considering that typical Reynolds numbers involved are of order of 1000 and therefore that viscous forces are negligible compared with convective ones. This leads to the Bernoulli law:

$$p + \frac{1}{2}\rho v^2 = \text{constant} \qquad (2)$$

where $p$ is the local pressure and $\rho$ the (constant) air density.

Equations (1) and (2) must be corrected in order to take into account a spectacular viscous effect: flow separation. Indeed, it is expected that the strongest pressure losses are due to the phenomenon of flow separation at the outlet of the constriction. This phenomenon is due to the presence of a strong adverse pressure gradient that causes the flow to decelerate so rapidly that it separates from the walls to form a free jet (see right part of figure 3). Very strong pressure losses, due to the appearance of turbulence downstream of the constriction, are associated with flow separation. As a matter of fact, the pressure recovery past the flow separation point is so small that it can in general be neglected. In the following, it is assumed that the flow separates from the walls of the constriction at the point where the area reaches 1.2 times the minimum area $A_0$ (see figure 3). This approximated value was empirically proposed and constitutes an acceptable approximation of the phenomena [10].

To summarize, for a given pressure drop ($p1$-$p2$), and for a given geometry of the constriction, the volume flow velocity $\Phi$ is:

$$\phi = A_s \sqrt{\frac{2(p_1 - p_2)}{\rho}} = 1.2 A_0 \sqrt{\frac{2(p_1 - p_2)}{\rho}} \qquad (3)$$

and the pressure distribution $p(x)$ within the constriction is predicted by:

$$p(x) = p_1 + \frac{1}{2}\rho\phi^2 \left( \frac{1}{A_1^2} - \frac{1}{A(x)^2} \right) \qquad (4)$$

where $A(x)$ is the transversal area at the $x$ abscissa (figure 3).

Therefore, the force exerted by the airflow onto the walls of the constriction can be computed by integrating the pressure along the *x* axis up to the flow separation point. This force induces a deformation of the upper airways soft tissues, thus modifying the airways geometry, and therefore changing the pressure distribution along the airways.

## 5. COUPLING THE AIRFLOW WITH THE VELUM MODEL

The coupling between the airflow pressure forces computation and the deformations of the FE model of the velum was iteratively processed. An adaptive Runge Kutta algorithm was used to solve the dynamical equations that govern the deformations of the velum. At each integration time step of the algorithm, the new deformed geometry of the velum is used to calculate the pressure forces distribution along the constriction. This new pressure force distribution is then defined as new boundary conditions for the Finite Element computation of the velum deformations.

As a first very qualitative approximation of the respiratory cycle, the pressure drop ($p_1$-$p_2$) was taken as a sinusoidal: $p_1$-$p_2 = p_{max}*sin(4*\pi*t)$. Figure 4 shows simulations of the airflow / velum interactions for a half-period pressure drop command with an 800 Pa maximal value. A clear reduction of the constriction can be observed, thus simulating a hypopnea [11].

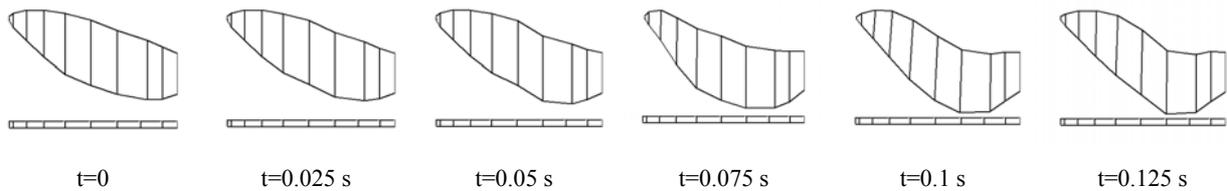

| t=0 | t=0.025 s | t=0.05 s | t=0.075 s | t=0.1 s | t=0.125 s |

Figure 4: FE model of the velum coupled with the airflow: from initial (left) to final (right) positions.

This decrease in the size of the constriction is associated with a limitation of the volume flow velocity. Left part of figure 5 plots this phenomenon, and shows also that this limitation can be avoid if the stiffness of the velum (the Young modulus value in our FE model) is increased. This point is qualitatively coherent with the clinical practice of uvuloplasty that consist in burning the velum in order to stiff the velar tissues. Similarly, the right part of figure 5 shows how a decrease in the size of the constriction can increase the volume flow velocity limitation.

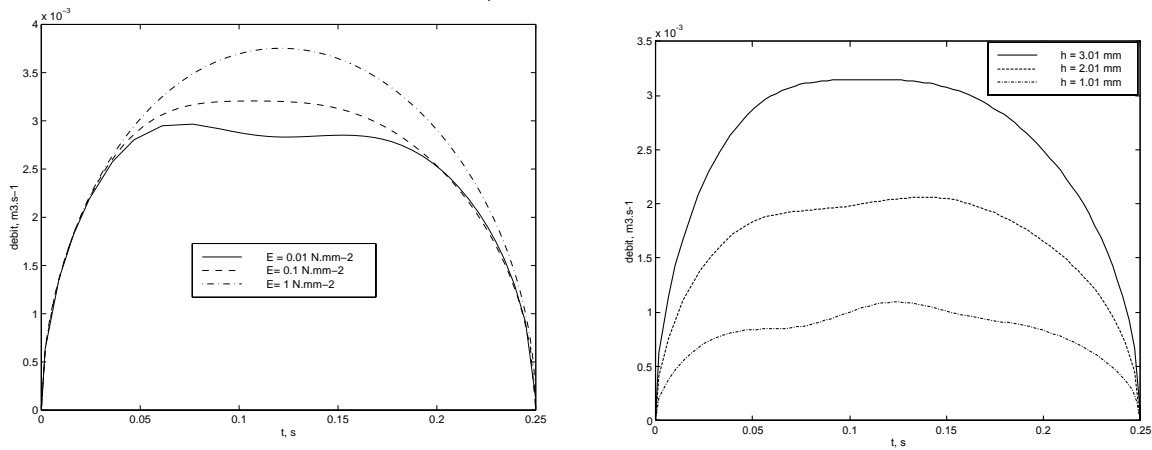

Figure 5: Simulated limitation of the volume flow velocity:
influence of the velum stiffness (left) and the constriction size (right).

## 6. EXPERIMENTAL SETUP

Our investigations are based on an experimental exploitation of a setup specially designed using classical dimensions obtained through in-vivo data acquisition. Although in vivo data serve as a reference for model development, they are highly non reproducible and don't allow quantitative validation of theory, as many parameters are not controlled, or even not reachable. This underscores the interest of using an experimental setup where most of the parameters are under control.

In order to investigate precisely the interaction phenomenon, the elastic characteristics of the upper airway soft tissues must be reproduced with sufficient realism. To fulfill this requirement, the soft tissues were modeled in a first approximation by a latex cylinder filled with water under pressure. This latex is assumed to represent to some extent the rounded backward part of the tongue, also responsible for a collapse during SAS. The latex is placed inside a squared rigid pipe (figure 6) that represents the larynx. It is assumed that larynx walls are rigid compared to the tongue elasticity. Extremities of the latex are fixed and glued to the pipe with silicone. Pressure measurements can be performed at different positions : before (*Psub*), after and at constriction (*Pgu*, *Pgd*). Complementary velocity measurements can also be made using a hot film (TSI, model 1210). Various parameters can also be controlled: air flow conditions in order to simulate different aerodynamic conditions, constriction height *hc*.

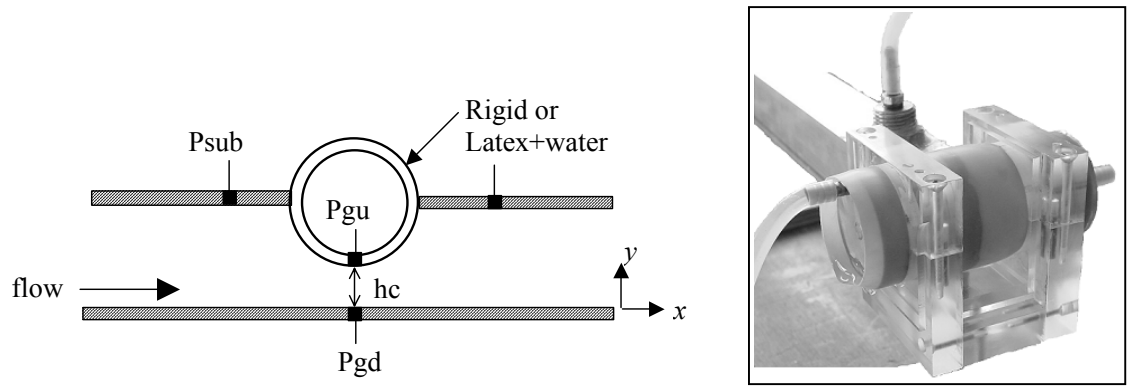

Figure 6: Schematic view and photography of the experimental setup
used to study airflow/tongue interaction.

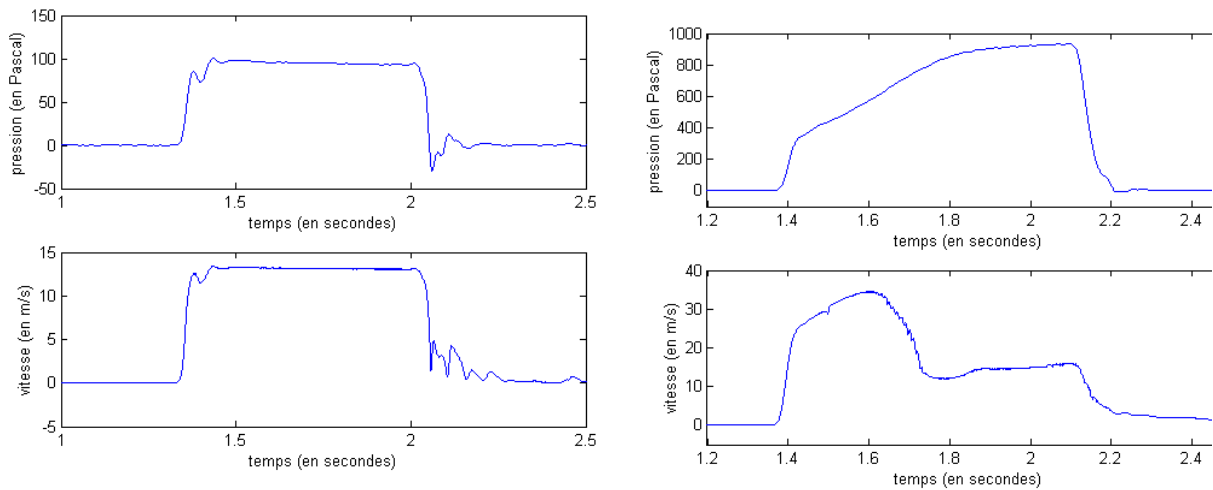

Figure 7: Pressure drop (top) and volume flow velocity (down) measured for two conditions:
    *Left*: low pressure drop; normal conditions
    *Right*: high pressure drop; pathological conditions with sleep hypopnea.

Figure 7 plots the pressure drop and volume flow velocity values measured at the constriction. When the pressure drop command has a low value (low respiratory effort), the volume flow velocity profile is similar to the pressure drop temporal evolution (figure 7, left). On the contrary, when the pressure drop command increases (close to 1000 Pa), the latex tends to collapse and the area of the constriction decreases. A limitation of the volume flow velocity is then observed (right lower part of figure 7).

## 7. DISCUSSION

Despite the limitations of our modeling hypotheses, preliminary interesting simulations were carried out. Indeed, the airflow model coupled with a 2D Finite Element model of the velum provides a decrease of the size in the constriction area with the increase of the pressure drop. This result is well known by the clinicians and is described as the airflow limitation phenomenon (hypopnea). Moreover, this phenomenon is observed in the *in vitro* experimental set up that we have developed. If they differ from a quantitative point of view, figure 5 and figure 7 (right lower part) both illustrate the same phenomenon: the volume flow velocity limitation. Finally, it is also interesting to note that an increase of the velum stiffness (modeled with an increase of the Young modulus value) or an increase in the size of the velo-pharyngeal constriction both tend to limit the hypo-apnea syndrome. Those results are consistent with some surgery techniques that try to modify mechanical properties of the velum (by burning tissues, thus increasing their stiffness) or try to have a more global and progressive action on the entire upper airways (in order to increase the size of the upper-airway constriction).